\documentclass[a4paper]{spie}  
\usepackage[]{graphicx}
\usepackage{upgreek}
\usepackage{booktabs} 
\usepackage{geometry}
\usepackage{textcomp}

\title{The digital data processing concepts of the LOFT mission} 

\author{C. Tenzer\supit{a}, A. Argan\supit{b}, A. Cros\supit{c}, Y. Favre\supit{d}, M. Gschwender\supit{a}, F. Jetter\supit{a},  A. Santangelo\supit{a},\\S. Schanne\supit{e}, P. Smith\supit{f}, S. Suchy\supit{a},  P. Uter\supit{a}, D. Walton\supit{f}, H. Wende\supit{a}
\skiplinehalf
\supit{a}Institut f\"ur Astronomie und Astrophysik, Sand 1, 72076 T\"ubingen, Germany\\
\supit{b}INAF-IAPS-Roma via Fosso del Cavaliere, 100, 00133, Rome, Italy\\
\supit{c}Institut de Recherche en Astrophysique et Planetologie, IRAP, 9 Avenue du Colonel Roche, BP44346, 31028, Toulouse, 
France\\
\supit{d}DPNC, Geneva University, Quai Ernest-Ansermet 24, CH-1211,Geneva, Switzerland\\
\supit{e}CEA/DSM/Irfu/Service d'Astrophysique, 91191 Gif-sur-Yvette Cedex, France\\
\supit{f}Mullard Space Science Laboratory, UCL, Holmbury St Mary, Dorking, Surrey, RH56NT,UK
}

\authorinfo{Further author information: (Send correspondence to C. Tenzer)\\
D. Maier.: E-mail: tenzer@astro.uni-tuebingen.de, Telephone: +49 (0)7071 29 75473}

\begin{document} 
  \maketitle 

\begin{abstract}
The Large Observatory for X-ray Timing (LOFT) is one of the five mission candidates that were considered by ESA for an M3 mission (with a launch opportunity in 2022 - 2024). LOFT features two instruments: the Large Area Detector (LAD) and the Wide Field Monitor (WFM). The LAD is a 10 m$^2$-class instrument with approximately 15 times the collecting area of the largest timing mission so far (RXTE) for the first time combined with CCD-class spectral resolution. The WFM will continuously monitor the sky and recognise changes in source states, detect transient and bursting phenomena and will allow  the mission to respond to this.

Observing the brightest X-ray sources with the effective area of the LAD leads to enormous data rates that need to be processed on several levels, filtered and compressed in real-time already on board. The WFM data processing on the other hand puts rather low constraints on the data rate but requires algorithms to find the photon interaction location on the detector and then to deconvolve the detector image in order to obtain the sky coordinates of observed transient sources. In the following, we want to give an overview of the data handling concepts that were developed during the study phase.
\end{abstract}


\section{INTRODUCTION}
\label{sec:introduction}

The Large Observatory for X-ray Timing (LOFT)\cite{LOFT} is one of the five mission candidates that were considered by ESA as an M3 mission (with a launch opportunity in 2022 - 2024) and has been studied during an extensive assessment phase in the previous three years. It is specifically designed to perform fast X-ray timing measurements and thus probe the state of matter near black holes and neutron stars. 

LOFT features two instruments: the Large Area Detector (LAD) is a 10 m$^2$-class detector with about 15 times the collecting area of the best past timing missions, which holds the capability to revolutionize the studies of variability from X-ray sources on the time scale of milliseconds. It will operate in the energy range 2 - 30 keV (up to 80 keV in extended mode) with good spectral resolution ($<$ 260 eV @ 6 keV FWHM) and a timing resolution of $10\,\upmu$s. Observing the brightest X-ray sources with the effective area of the LAD leads to enormous data rates that need to be processed, filtered and compressed already on board. The Wide Field Monitor (WFM) located on the top of the spacecraft is a coded mask instrument with the capability to simultaneously observe 1/3 of the sky in the same energy band as the LAD. The main purpose of the WFM is to detect transient and bursting sources for follow-up observations with the LAD, so the field of view is designed to have a maximum overlap with the sky accessible to LAD pointings. The Wide Field Monitor monitors a large fraction of the sky for changes in the state of targets of interest (especially NS and BHs, but AGNs as well) or discovery and localization of new sources. If a relevant transient is detected, the viewing direction of LOFT can be modified to observe this source with the LAD.

Although LOFT was not downselected for launch, during the assessment most of the trade-offs have been closed, leading to a robust and well documented design that will be reproposed in future ESA calls.

\section{Detectors and Front-End Electronics} 
\label{sec:DetectorsFEE}

\subsection{The Silicon Drift Detectors}
\label{SDDs}

The LOFT detectors are designed on the heritage of the Inner Tracking System of the ALICE experiment at the Large Hadron Collider\cite{Alice}. 
They are double-sided linear drift detectors, having an area of 120 $\times$ 72 mm$^2$ (which is the maximum size possible on a 6-inch wafer). The ALICE SDD naturally shares with all other SDDs a low anode capacitance of the order of 100 fF that allows for a very low noise contribution from the Front-End Electronics (FEE). The charge generated by the absorption of an X-ray photon is collected in the middle plane of the detector bulk and then drifted towards the read-out anodes on one edge of the detector by means of a negative potential with a steep gradient requiring an input HV of 1300 V. The electric field is sustained by a series of cathodes on both sides of the detector. The detectors will be paired with a collimator based on lead-glass micro-capillary plates to constrain the FOV to 1 degree.

The drift concept combines the large collection area with good spectroscopic performance in the LAD instrument: On each detector half there are 292 drift cathodes with a pitch of $120\,\upmu$m and only 112 readout anodes (thus low power requirements, but still CCD-type energy resolution) with a pitch of $970\,\upmu$m. The sensitive-to-total-area ratio is 87$\,\%$. Aside from the integrated dividers there are two triangular areas constituted by the guard p+ cathodes. The detector is planned to work at a drift field of 360 V/cm, entailing a maximum drift time of about $7\,\upmu$s at $20\,^{\circ}$C, which is reduced to about $5\,\upmu$s at $-20\,^{\circ}$C due to the higher electron mobility at lower temperatures.

The WFM design is based on the same detector type and the general working principle of the SDD detectors is the same as for the LAD, but the detectors of the WFM are designed to optimise imaging properties when combined with the coded mask (224 read-out anodes per detector half and square detector dimensions).

The final detector prototype design, delivered by the end of 2013, has been manufactured by Fondazione Bruno Kessler in Trento, Italy, under design by INFN-Trieste. The detector production was carried out using 6-inch diameter floating-zone (FZ) Silicon wafers, a resistivity of approximately $9\,$k$\Omega\times$cm and a thickness of $450\,\upmu$m. 

\subsection{The Front-End Electronics}
\label{FEE}

\begin{figure}[h]
\centering
\includegraphics[width = .5\linewidth]{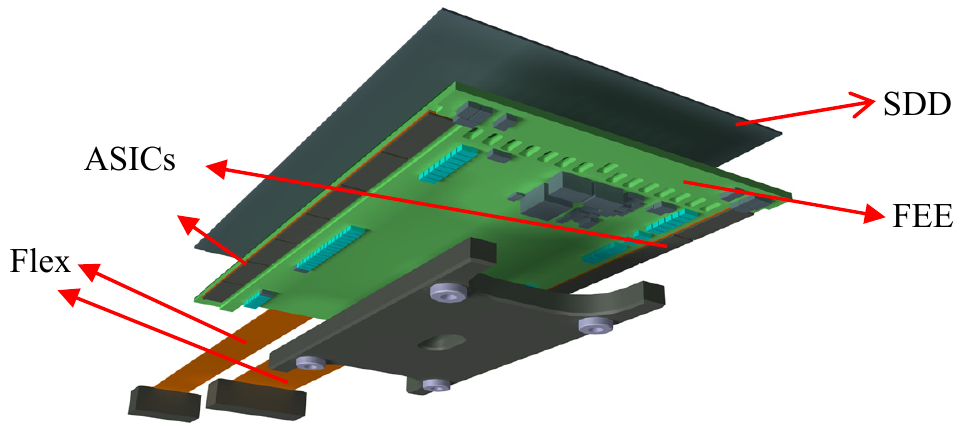}\includegraphics[width = 0.5\linewidth]{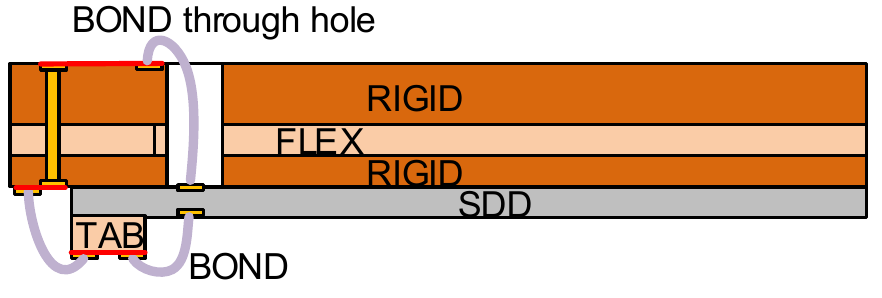}
\caption{Left: CAD design of the detector and its front end electronics. The SDD is shown at the top and is glued to a FEE-PCB. This PCB holds the 14 ASICs which are bonded directly to the SDD anodes. Also seen are two flexible leads (power and signals) to the MBEE. Right: FEE-SDD bonding principle for the LAD}
\label{WaferSideAndFEE}
\end{figure}

The Front-End Electronics (FEE) PCB is made from a rigid-flex technology. No through holes components are possible because the SDD is glued on the bottom side of the FEE PCB. The SDD is a double side detector with its anodes located on the top and bonded directly to the ASIC inputs with a short wire bond for capacitance reduction (i.e. noise limitation). The bottom cathodes of the SDDs are first bonded to an intermediate flex tab glued on the bottom of the PCB and the tab extremity is then bonded to the PCB. This double bond mechanism is used for wire length reduction. The top cathodes of the SDD are located under the PCB and holes are drilled into it in order to allow the bonding (see Figure \ref{WaferSideAndFEE}).

The front-end read-out of the LAD and WFM detectors will be based on mixed-signal ASIC technology\cite{cros}. As the LAD and WFM use the same type of detector, the ASIC design is the same, with the only differences being related to the different anode pitch, and consequently stray capacitance and leakage current. Adapting the LAD ASIC pre-amplifier to the lower leakage current and capacitance of the smaller WFM anodes allows to directly meet the noise requirements. The FEE is separated in two symmetrical rows of seven ASICs; each ASIC has its input channels (16 for the LAD and 32 for the WFM) directly connected to the SDD anodes. A total of about 28.000 ASICs are needed to build LOFT.

The very large number of read out channels implies the use of mixed signal ASICs with low power consumption and good performance, whose feasibility has been already demonstrated. The ASICs amplify and digitise the anode charge pulses resulting from X-ray events (650 uW per channel, 17e- rms per read-out channel). The ASIC for LOFT is a new development carried out by Dolphin Integration. The development of this so-called Sirius ASIC was kicked-off on May 2012. The first prototype, aimed at demonstrating the most critical parameters (low- noise and low-power on the analogue section), is an 8-channel device, with 4 channels with $145\,\upmu$m pitch and 4 channels with $970\,\upmu$m pitch. Each channel includes the key analogue functions: charge pre-amplifier, shaper amplifier, peak detection and hold, with their associated electronics and test input. The shaper is an RC-CR$^2$ type, with selectable shaping time between 1 and $4\,\upmu$s. The ASIC is implemented using TSMC MS/RF $0.18\,\upmu$m CMOS technology.

\begin{figure}[h]
\centering
\includegraphics[width = .8\linewidth]{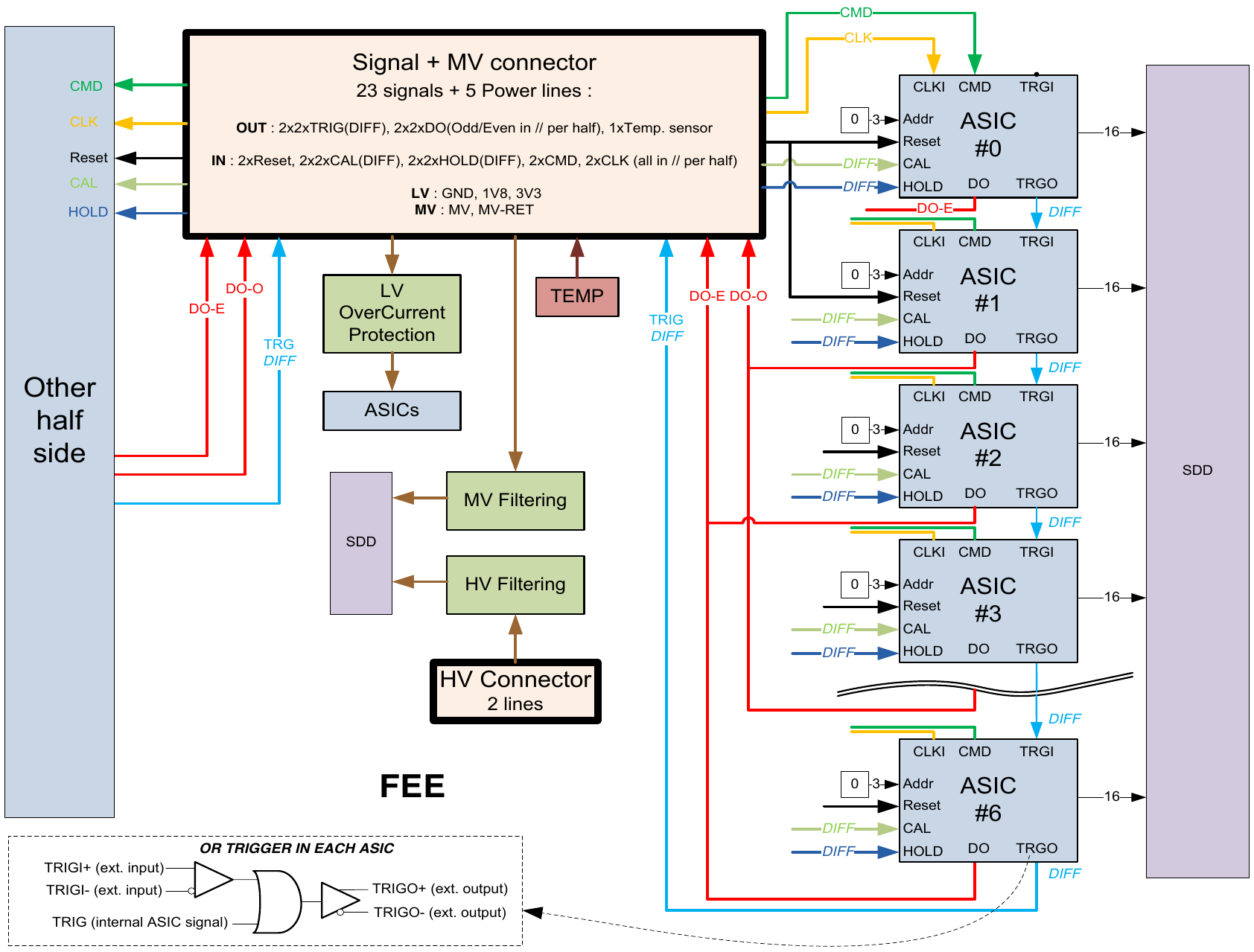}
\caption{Schematic drawing of the interface between the FEE and (M)BEE showing the signals to and from the ASICs.}
\label{interface}
\end{figure}

Each ASIC on a detector half is chained by a differential trigger line allowing triggers to be propagated from the first to the last ASIC and to the Module Back-End Electronics (MBEE) where the digital processing takes place. The MBEE is also connected to the trigger input of the 1st ASIC and can force a trigger to all ASICs in order to perform noise measurements. Each ASIC has also an individual hardwired address allowing its individual access through the common command (CMD) and clock (CLK) differential signals where individual ASIC address (0 to 6) or broadcast (7 i.e. all ASICs) can be set within the command frame. Reset and Hold signals are also common to each row. Data outputs for the digitised amplitudes and trigger details are divided into 2 differential pairs of signal per FEE side: Odd and Even DOUT. This separation allows a higher bandwidth when all ASICs send their data especially when a trigger occurs on channels on two adjacent ASICs: in this case, a single DOUT frame can be processed simultaneously on the 2 ASICs (see Figure \ref{interface}).
The 2 ASIC rows are connected through Hyperstac connectors located at the end of the flex extensions: one 40 pins connector for low voltage digital and power supply signals and another 4 pins connector for the SDD high voltage and medium voltage signals. Hyperstac connectors have been chosen for their simplicity of connection and cost: no soldering operation is required during manufacturing. High voltage can also be passed through it thanks to a specific design for LOFT done by the manufacturer with appropriated insulation distances.

\section{The Large Area Detector} 
\label{sec:LAD}

The required large area of the instrument will be achieved by a modular design (called LAD modules). Each module will have a set of 4 $\times$ 4 detectors and 4 $\times$ 4 collimators including the electronics needed for biasing and read-out. It will interface the rest of the instrument through a hierarchical digital and power interface only. The LAD modules will be mounted on deployable panels of the satellite. Depending on the satellite design there will be either two, five or six panels and the number of LAD modules also differs slightly (124-126 modules). To meet the effective area requirement, at least 120 modules are needed. 

Each photon is detected in an SDD layer 450 microns thick, and the cloud of electrons liberated then drifts towards the anodes. The dynamic range of the read-out ASICs allows operation in the energy range 2-80 keV. Dead time of the instrument (induced only by the front-end readout and digitisation process) is accurately controlled via several counters as this is very important for a timing instrument. The field of view of the LAD will be limited to $<$1 degree by X-ray collimators. These are micro-channel plates based on developments for the BepiColombo mission\cite{BepiColombo} and consist of a 5 mm thick sheet of Lead glass with a large number of square pores with $83\,\upmu$m pore width and $16\,\upmu$m wall thickness, giving an open area ratio of 70$\,\%$. The stopping power of Pb in the glass over the large number of walls that off-axis photons need to cross is effective in collimating X-rays below 30 keV. At the back side of each module there will be a radiator (for the passive cooling of the detector) and a shielding will be applied in addition to reduce the background. 

\begin{figure}[h]
\centering
\includegraphics[width = .7\linewidth]{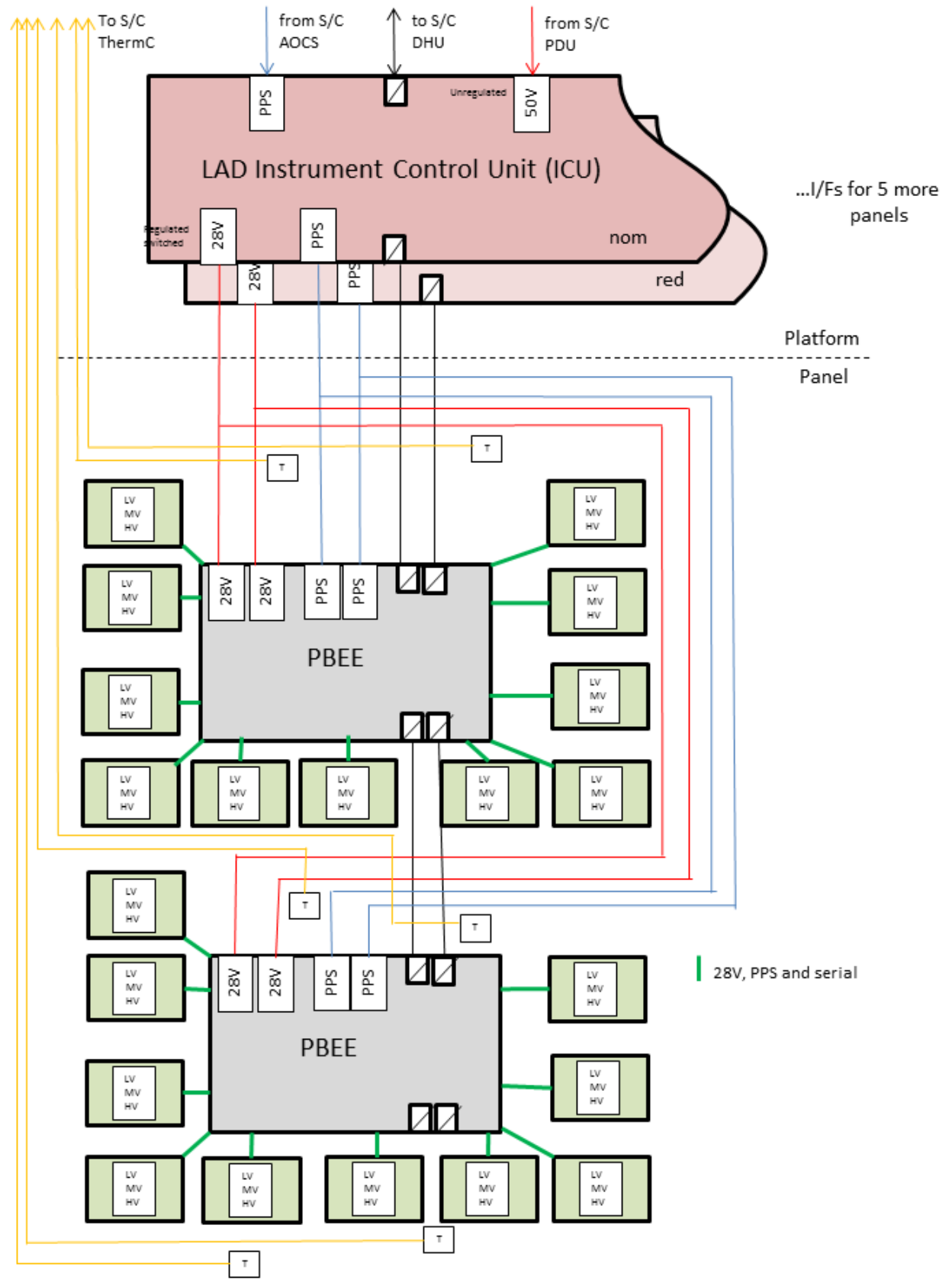}
\caption{Electrical architecture for the LAD. Shown are the connections for power and time distribution as well as data handling between MBEEs (green), PBEEs (grey) and the ICU (red) for one LAD panel.}
\label{ladarchitecture}
\end{figure}

The consortium design for the LAD configuration is based on six detector panels (approximately 1 $\times$ $3\,$m$^2$ each), connected by hinges to an optical bench at the top of a tower\cite{zane}. The panels will be tiled with 2016 detectors in total, electrically and mechanically organised in the modules. The read-out electronics employs a hierarchical approach and is designed as follows: the FEEs of the 16 detectors in a module converge into a single Module Back-End Electronics (MBEE). Two Panel Back-End Electronics (PBEE) for each panel are in charge of interfacing the MBEEs. The data collected by all PBEEs are passed on to the Detector Handling Unit (DHU) which is part of the Instrument Control Unit (ICU). The DHU collects and compresses data from all PBEEs in a mass memory for ground transmission (see Figure \ref{ladarchitecture}).

\subsection{Trigger Scheme}
\label{sec:trigger}

When the collected charge exceeds a programmable threshold in one of the ASIC channels in the Front-End Electronics, a trigger signal is forwarded to the MBEE where a time tag (based on a centrally synchronised internal clock count) is instantaneously generated. The trigger is also propagated directly from one ASIC to all the ASICs on the respective detector half and they all freeze their current signal. The MBEE requests a trigger map (list of anodes with signal above an adjustable threshold) from all the ASICs on this detector half and validates if only one or two adjacent anodes triggered. If the event passes this selection criteria, the "A/D conversion" command will be sent from the MBEE to the respective ASICs and the conversion will be carried out (inside the ASIC), providing an 11-bit output per anode.

\subsection{Module Back-End Electronics}
\label{sec:mbee}

Following the A/D conversion, the MBEE processing pipeline will be activated. The saved time tag will be added to the event package later at the end of the processing pipeline. The main processing functions of the MBEEs are time tagging, trigger validation and filtering, pedestal and common noise subtraction, energy reconstruction, event threshold application and collection of housekeeping data. Each MBEE consists of two PCBs and is designed to process events within a pipeline structure that handles the events from one side of a detector (7 ASICs). This pipeline is initiated 16 times within the two MBEE FPGAs to allow processing of data from all detectors simultaneously. The pipeline is designed such that the processing time within each step is the same and shorter than the A/D conversion time of a possible following event. In this way, the data processing in the MBEE does not inflict additional dead-time and the pipeline is always ready for the next event.

The first step within each pipeline is the pedestal subtraction where a set of pedestal values (one for each anode) stored within the MBEE are subtracted from the measured signal values. Then, the common noise (CN) is calculated and also subtracted. The CN is a noise component common to all the channels connected to the same ASIC and basically an undesired baseline shift. It is composed of two effects: the CN produced by the detector and the CN introduced by the ASIC. Thus, the CN will be calculated independently for each ASIC considering only the channels not affected by the charge cloud. A median algorithm will be used for the calculation. The next step of the pipeline concerns the on-board energy reconstruction. The procedure will be composed of the following steps: 1) Gain Correction: the reconstructed energy of the event is the sum of the n event channels, each one multiplied by an individual gain factor that is kept in a lookup table which in the same fashion as the pedestal table can be recalculated or uploaded from ground. 2) Temperature Correction: The gain calibration will be performed at a fixed temperature and a gain variability factor of 0.1\% per degree is assumed. A linear correction $E=E\cdot(1-C\cdot(T-T_0))$ is applied to the energy value of the event. For a realistic 5 degrees for $T-T_0$, we obtain a correction factor C of 150 eV at 30 keV which is not negligible. 3) Upper Threshold Rejection: If the reconstructed energy exceeds an upper threshold, the event is rejected. 4) Energy Scaling: The last step of the on-board energy reconstruction generates the final 9 bit energy word according to a non-linear function as follows: from 2 keV to 30 keV: 60 eV per digit and from 30 keV to 80 keV: 2 keV per digit.

The TM/TC interface between MBEE and PBEE is using the SpaceWire hardware (cable and connectors) but does implement a custom, SPI-like data transfer with less overhead running at 10 MHz.

\subsection{Panel Back-End Electronics}
\label{sec:pbee}

The Panel Back-End Electronics (PBEE) handles all events from the individual modules of one of the six detector panels. It is the heart of the data acquisition and signal processing, located between the individual modules and the DHU on the satellite bus. The main tasks of the PBEE are: interfacing the MBEEs, Collecting and buffering the event packets, differential time assignment, transferring the data to the Instrument Control Unit, collection of HK data and creation of HK packets.

In addition, depending on a selectable observation mode for very bright sources, the PBEEs have the ability to integrate the data from the MBEEs over a certain amount of time into configurable spectra, thus dropping the actual event-by-event data in order to comply with the available telemetry  bandwidth. The interface between PBEE and DHU is a fully compliant SpaceWire interface, runnnig at 100 MHz. 

There is no redundancy at the PBEE level as the loss of one of the six panels would not compromise the scientific goals of this mission. In the case of a S/C design with fewer Detector Panels, segmentation of the PBEE is necessary in order to maintain an appropriate level of graceful degradation in the event of a single PBEE failure.

\subsection{Instrument Control Unit}
\label{sec:icu}

The Instrument Control Unit forms the ÔmajorÕ controlling element of the LAD instrument. It provides the interface to the spacecraft On-Board Data Handling (OBDH) and also control of all LOFT instrument sub-systems via SpaceWire. The ICU box consists of three components: the Data Handling Unit (DHU), the mass memory and the Power Distribution Unit (PDU). The tasks performed at the level of the ICU involve telecommand execution and distribution, access to mass memory, time distribution,
data processing and compression, HK collection, instrument health monitoring and calibration tasks.
The ICU box is provided twice for cold redundancy. At its heart is the Data Handling Unit where the scientific data is compiled, processed and also compressed with a lossless compression algorithm. A LEON III processor was chosen for the task to run the onboard software due to its additional flexibility when compared to a hardware only architecture, i.e. a state-machine. The DHU collects the data from the PBEEs via a SpaceWire interface PCB, performs the selection and formatting tasks depending on the selected observation mode and commits the data to the mass memory in a compressed format. Besides the routing and execution of telecommands, the ICU can run macro commands for several tasks, including on-board calibration and the individual sequenced switching of the MBEEs in the engineering modes.

The LAD on-board software runs on the LEON III and its main functions are instrument control and monitoring as well as science data processing and formatting. Using software will allow the instrument to have the complex functionality that it requires to allow itself to be updated and work around problems automatically and with input from the ground. Instrument control will be possible through the software via telecommands from the ground (e.g., power on and off, set-up of ASICs and FPGAs, loading parameters for processing/on-board calibration, investigations) and autonomously on-board (e.g., mode switching, FDIR and diagnostic data collection).The software will implement the standard ECSS-style PUS service telecommand packets for housekeeping, memory maintenance, monitoring etc. and some standard telemetry packets for command acceptance, housekeeping, event reporting, memory management, time management, science data etc.
The software will be able to send setup information to the hierarchy of processing elements and receive and process the housekeeping data coming back, simplifying these data in a configurable way for a lower rate transmission to the ground.

The software will be written in C using RTEMS, a real-time executive, which will schedule the software tasks, each at different priorities, communicating with each other with messages and events. The software will be organised as two separate sub-systems: \textit{Basi}c software, stored in very reliable PROMs, would have enough functionality to perform basic health management and memory management: to receive, store and execute patches or new software. \textit{Operational} software, stored in EEPROM, would have the functionality of the \textit{Basic} software and also the full science capabilities. In this way, new software can be loaded to the instrument without overwriting or corrupting the original PROM code. If there is any problem with the software interface to the spacecraft, a reboot or power off/on of the instrument will reset the software into the well-tested \textit{Basic} mode which does not produce science data, thereby initialising the instrument back to a well-defined mode. As the software has to operate in a remote space environment, it will be written to be robust against errors, to report as much information as possible on any problems encountered and progress made (to help investigations) and perform any operations required by EDAC/memory scrubbing. Error messages will be carefully controlled so that the spacecraft is not flooded with messages that may have a common cause. A software-controlled hardware watchdog will be implemented.

The interfaces between the software and the rest of the instrument/satellite will be as clean as possible with the processor and software taking over the processing of the events and diagnostic data at the point where the data streams from the panels are joined. Interaction with the Spacecraft will be through SpaceWire (with the option of CAN).
\subsection{Power Distribution}
\label{sec:power}

Power will be distributed in the form of 28 Volts from the S/C PDU via the ICU and the PBEEs to the individual MBEEs. All necessary lower voltages are generated at the respective level (ICU, PBEE, MBEE) by local power converter boards. This design allows for a minimum of harness over the panel hinge (where the bending radius is important) and from the PBEEs to the MBEEs (where the overall weight of the harness is strongly affected by the high multiplicity of the modules). The high voltage necessary to operate the detectors is also generated by a dedicated HV-daughterboard in the individual modules.

\subsection{Event Rates and Dead-Time}
\label{sec:rates}

Time information will be received from the GPS and distributed via the PBEEs to the MBEEs with a synchronised 1 MHz clock line and a Pulse-Per-Second signal. The MBEEs will monitor their internal time counters with respect to these lines in case re-synchronisation is needed. After an event in one of the detectors, all ASICs on the respective detector half will freeze their signals and this half of the detector will be dead for the time the data processing takes. This action is chosen deliberately in order not to minimise but to homogenise the dead area of the LAD at a given time in order to facilitate the flux reconstruction.

There are three different event processing scenarios to be accounted for when calculating the dead-time of the instrument:
1) An event is discarded due to an invalid trigger pattern (e.g. events caused not by photons but by minimal ionizing particles). The respective detector half is rendered insensitive for about $<10\,\upmu$s until the decision to discard the event has been taken and communicated by the MBEE.
2) The regular event-processing dead-time (mostly A/D conversion). The respective detector half is rendered insensitive for about $100\,\upmu$s until the event data has been digitised and communicated to the MBEE.
3) The regular event-processing dead-time (mostly A/D conversion) for an event that is later on discarded in the processing for being outside of the energy range. It is the same as the dead-time for a regular event as the processing is not inflicting additional dead-time on the detectors.
Counters in the MBEE HK keep track of the rates of these three event types.

All later data processing and transmission steps are designed to handle the data in real-time so the current bottleneck for the dead-time remains the time it takes to digitise and transmit the event from the FEE.

The standard candle of X-ray astronomy, the Crab Nebula, is expected to provide a count rate as high as 240,000 cts/s. 
The amount of data generated from these events can not be directly transmitted to ground as the available bandwidth is too small.
The LAD telemetry budget is almost solely driven by the LAD event-by-event data (24 bits per event). At a source strength equivalent to 1 Crab this corresponds to 2460 kbps for this data. Adding the scientific rate meters and housekeeping increases this to 3740 kbps. Lossless data compression will be applied in addition to all observational data stored in the mass memory. Taking into account a compression factor of 1.77 (verified by software tests) this gives a total rate of 2040 kbps. For stronger sources (above 0.5 Crab) the options for additional rebinning in the PBEEs automatically after a definable time (this will stop the recording of event-by-event data and directly create spectra that are configurable in time and spectral resolution by the observer) and onboard storage of event data up to 15 Crab for 300 minutes is foreseen. The system is designed to handle a maximum flux of up to 40 Crab for a short time in order to record bursting or flaring activity. These data can then be transmitted during the following orbits or during the interleaved observations of fainter sources.

\section{The Wide Field Monitor} 
\label{sec:WFM}

The Wide Field Monitor (WFM) is an instrument with the capability to simultaneously observe 1/3 of the sky (44\% at zero response, and 33\% at 20\% of peak response) in the same energy band as the LAD. The main purpose of the WFM is to detect sources for follow-up observations with the LAD, so the field of view is designed to have a maximum overlap with the sky accessible to LAD pointings. These sources are new transients, as well as known sources undergoing spectral state changes.
The WFM is a coded mask imaging instrument with a point source location accuracy of 1 arc minute, matching the required on-target pointing accuracy for LAD observations. 

The WFM is based on a modular design with ten identical cameras\cite{brandt}. It employs the same basic detector technology as the LAD, but with a read-out scheme that enables 2D position determination of the X-ray interactions required for the imaging. However, the detector position resolution in one direction is much finer ($<100\,\upmu$m) than in the other direction ($<8\,$mm). This is reflected in the elongated point spread function of the de-convolved sky images. Therefore, the 10 WFM cameras are grouped into 5 camera pairs with two co-aligned cameras rotated 90$\,^{\circ}$ around their viewing direction. By observing simultaneously the same sky region, one can derive the precise 2D position of the sources, by intersecting the two orthogonal 1D projections. Each camera holds a detector tray with 4 SDDs, 4 Front-End Electronics, and the Back-End Electronics (BEE).	

\begin{figure}[h]
\centering
\includegraphics[width = .9\linewidth]{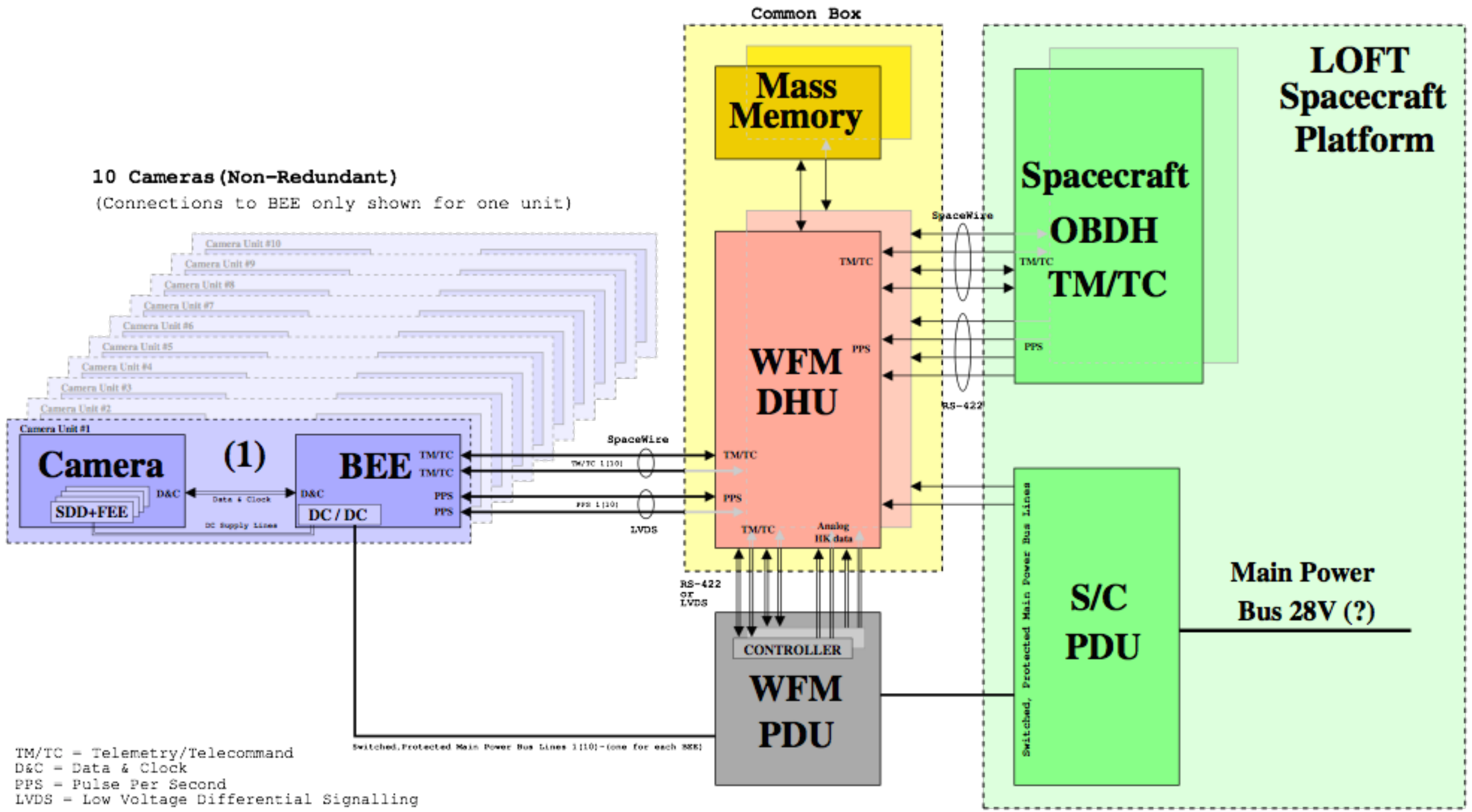}
\caption{The WFM electrical architecture showing the 10 cameras and their Back-End Electronics on the left as well as the power and data connections to the Instrument Control Unit holding the DHU, PDU and the Mass Memory.}
\label{WFMArchitecture}
\end{figure}

The electrical architecture of the LOFT WFM is shown in the schematic \ref{WFMArchitecture} and encompasses all electrical subsystems of the WFM and their interfaces with the LOFT spacecraft platform. Electrically, the central part of the WFM is the Instrument Control Unit (ICU), which comprises the Data Handling Unit (DHU) with Mass Memory and a Power Distribution Unit (PDU). The ICU consists of a main and a redundant unit which are housed in two separate boxes as shown in the schematic. The LOFT Burst On-board Trigger (LBOT) functionality is implemented as a part of the DHU.
The DHU is directly attached to the LOFT Spacecraft On-board Data Handling (OBDH) System. The electrical interface is assumed to be SpaceWire. The SpaceWire connections are cross-strapped between the main and redundant DHU and the main and redundant OBDH. The same goes for the Pulse-Per-Second (PPS) lines, which are assumed to be RS-422 interfaces. Mass memory for storing WFM data until downlink is an integral part of the DHU. The electrical interface is SpaceWire, and each BEE is connected to both the main and redundant DHU, both for control, data, monitoring. The bus power is routed through the WFM PDU providing ON/OFF switching and protection capability. As both main and redundant bus power is supplied to every BEE, an OR-type blocking diode circuit is required at each BEE power input as shown in the schematic.

\subsection{WFM Front-End Electronics}
\label{sec:WFMFEE}

The selected base material for the Front End Electronics is Aluminium Nitride (AlN), applied as ceramic thick film substrate. It has several advantages with respect to FR4 and polyimide like better thermal conductance, lower CTE and better wire bonding capabilities.

Because of the fine pitch of the anodes on the SDD, it is not feasible to have the ASICs packaged. Therefore, they will be placed as bare dies on the FEE. The two rows of ASICs will be glued to the FEE very close to the two edges near the anodes of the SDD. Aluminum wire bonds will used to connect the ASICs to the anodes and to the FEE. The bondings from anodes to ASIC inputs will be made as short as possible and thin ($17\,\upmu$m or $25\,\upmu$m) for low capacitance.
Most of the AlN substrate will have a thickness of 1.5$\,$mm, but at the edges it will be 0.5$\,$mm for minimum wire bond length.

Connection of the low voltage supplies and the digital signals between BEE and FEE will be made via a flexprint. At both ends, the flexprint will have a rigid part of PCB. At the FEE side, the rigid part will be glued to the FEE and connected by means of wire bondings. At the BEE side this will be connected to a D-type connector.
The high and medium voltage supplies from the BEE will be fed towards the FEE via two coaxial HV cables. At the BEE side these cables will be equipped with Reynolds connectors and at the FEE side they will directly be mounted to the PCB. On the FEE these voltages will be filtered and the medium voltage will be divided before they are connected to the SDD.
Instead of the earlier proposed wrap-around cable for interconnection of high voltages between both sides of the SDD, additional strips (tabs) of Aluminium Nitride susbstrate will be glued on top of the SDD. The high and medium voltages on both FEE sides will be interconnected with metal clamps and from the tabs to the SDD the connection will be made with wire bonds.

\subsection{WFM Back-End Electronics}
\label{sec:BEE}

The WFM BEE differs significantly from the corresponding LAD unit (the Module Back-End Electronics), because of the absence of the PBEEs in the design (direct SpaceWire-Link required) and the need in the WFM to calculate the location of each photon in the detector plane in order to reconstruct the images. The electronics therefore has the task to derive the position of the interaction point with a precision of $\approx 20\,\upmu$m along the direction of the line of anodes and about 4$\,$mm in the drift direction. The most obvious method for the reconstruction of the event location from the energy corrected charge distribution along the anode is to fit the shape of the measured charge cloud with a Gaussian or more correctly an erf-function distribution, which determines the position of the peak and the width of the distribution, and which then can be used (together with the total charge signal) to reconstruct the drift distance. Other possibilities could be the implementation of a center-of-gravity method or lookup-tables.

The main functions of the BEE are: time tagging of the X-ray events, trigger selection, pedestal and common mode noise subtraction, determination of charge cloud centre and width (position in the fine and coarse direction) and
the reconstruction the total collected charge. The WFM BEE also takes care of collecting the HK data from the sensors placed inside the camera housing and controls the power supply board and heaters.
The BEE board is based on a VIRTEX 4 FPGA (XQR4VSX55) device driven by a 60 MHz system clock. The chip is radiation tolerant: TID $>$300 krad (Si) and Dose Rate Latch-up $>$107 Rad(Si)/s. All functionality will be implemented in the FPGA firmware. MDM-31 connectors will be used to connect the BEE box to the FEE for the digital interface.

\subsection{Reconstruction of the Interaction Location}
\label{sec:Positon}

The back-end electronics has the task to derive the position of the interaction point along the direction of the line of anodes and in the drift direction. The reconstruction of the event location as the center of gravity from the energy corrected charge distribution can be used to achieve a higher spatial resolution than the implemented anode pitch. The determination of the width of the charge cloud can be used to estimate the drift distance, and give a rough spatial resolution towards the middle of the detector (orthogonal to the anode direction). 

The most obvious method to determine the shape of the charge distribution is therefore, to fit the shape of the measured charge cloud with an erf-function distribution, which corresponds to the projection of the three-dimensional charge cloud onto the anode side. From the fit one determines the position of the peak and the width of the distribution, which then can be used (together with the total charge signal) to reconstruct the drift distance. This is done by a fit of the single channel values distribution with the function 
\begin{equation}\mathrm{Anode}_x = q\cdot[erf( (x+0.5-m)/s ) - erf( (x-0.5-m)/s)]\end{equation} 

with x = anode position, q = total charge and the fitting parameters m = center of distribution and s = width of distribution.
An alternative (simpler) approach is the usage of the weighted centroid method, where the centroid of the charge distribution is calculated by summing over the product of each anode and its position and then dividing by the total charge: 

\begin{equation}\mathrm{centroid} = \displaystyle\sum\limits_{i=0}^n (C_i \cdot x_i)/ \displaystyle\sum\limits_{i=0}^n(C_i).\end{equation} 
Assuming a Gaussian distribution, the width can then be calculated by using the following equation: \begin{equation}\mathrm{width} = ((1/\sqrt{2*Pi}) \cdot q / amplitude ) \cdot pitch.\end{equation}
In the final event packet that is passed on to the DHU, differential time is used instead of absolute time for the time tags in order to reduce the amount of data to transfer. To calculate the differential time, the events are first stored within a buffer, reordered if necessary and finally the time difference is calculated. If the evaluated differential time is greater than the upper limit of the codeable range, a dummy event will be generated. Every 100 ms an Absolute Time Event will be generated in order to stop errors from propagating through the differential time from event to event.

The event packet passed on to the ICU will also give the location of interaction in a global camera-coordinate-system instead of detector specific coordinates. This will allow for taking into account small rotational misalignments between the individual detectors and of course the respective detector offsets.
The TM/TC interface between BEE and ICU is according to the SpaceWire standard.

\subsection{WFM Instrument Control Unit}
\label{sec:ICU}

Electrically, the central part of the WFM is the Instrument Control Unit (ICU), which comprises the Data Handling Unit (DHU) with the Mass Memory and a Power Distribution Unit (PDU). The ICU will be based on the Virtex-5QV (XQR5VFX130) FPGA device driven by a 450 MHz system clock and provides $>$100 DMIPS when a modular, ESA-Qualified LEON3FT VHDL IP Core is implemented. The chip has ample radiation tolerance: TID $>$1Mrad (Si) and Dose Rate Latch-up $>$1010 Rad(Si)/s. The main functions of the WFM ICU are: interfacing the BEEs, TC and configuration handling, on-board time management, image integration, burst triggering.
The Virtex-5QV FPGA is configured by loading the configuration data from an external PROM into a radiation hardened RAM on the chip. If the PROM is replaced by an EEPROM or Flash memory, the FPGA will become in-flight reconfigurable, which has clear advantages. A total of 12 SpaceWire interfaces are needed and therefore a dedicated SpaceWire core is included.

An important part of the DHU is the LOFT Burst Onboard Trigger system (LBOT, see below) functionality. The LBOT is designed to detect and localize fast astronomical transient events in real-time (e.g. Gamma-Ray Bursts). It produces corresponding near real-time alert messages for dissemination via a VHF system to the world-wide scientific community interested in follow-up observations. The complex deconvolution process required to localize the source on the sky is foreseen to be implemented in software; to gain CPU power margin, this process could be sped-up by hardwired 2048 and 32 point, radix-2 FFT processors and a hardwired complex multiplier.

\subsection{Time Distribution}
\label{Time}

Time pulse distribution will be implemented via the SpaceWire interface outlined above using Time-Codes, which are part of the SpaceWire standard. They provide a means of synchronising units across a SpaceWire system to spacecraft time. The PPS input from the GPS receiver to the platform OBDH will be a prerequisite and SpaceWire is used for time-code distribution all the way from the OBDH to the BEE and a timing accuracy of $\pm$500 ns has to be maintained. The distribution of time-codes using the SpaceWire protocol gives rise to a time skew or latency across a network of Tskew = 14$\cdot$N/B, where N is the number of SpaceWire links traversed, and B is the link bit rate. The time skew is deterministic  - using e.g. a 25 Mbit/s network, a time skew of 560 ns will be incurred per link traversed. For the LOFT WFM, two links will be traversed from the OBDH to the BEE, thus incurring a time skew of 1.12$\,\upmu$s.
Jitter is also introduced at each link interface due to the variation in time spent waiting for the transmitter to finish transmitting the current character or control code. Jitter is non-deterministic. This variation can be anywhere from 0 to 10 clock periods, or 0 to 400 ns for a 25 Mbit/s SpaceWire link. This implies $\pm$200 ns of jitter with a uniform distribution over this interval. For two links the jitter will have a triangular distribution over a $\pm$400 ns interval (the convolution of two identical uniform distributions). In addition there is a small amount of jitter stemming from the unsynchronized clocks of the transmitter and the receiving SpaceWire interface. This implies a uniformly distributed jitter with a width on one clock period. Assuming a 25 MHz internal clock of the SpaceWire interface the jitter will be $\pm$20 ns.
In total for a 25 MHz SpaceWire bus, traversing two links, the total jitter would be less than $\pm$440 ns, in fulfillment of the science requirement. However satisfying the goal of 0.5$\,\upmu$s absolute statistical time accuracy, the clock frequency must be doubled to 50 MHz, leading to a jitter of less than $\pm$220 ns.

\subsection{LBOT}
\label{LBOT}

The LOFT Burst Onboard Trigger system (LBOT) is designed to detect and localise fast astronomical transient events in real-time (among which are Gamma-Ray Bursts)\cite{schanne}. The LBOT produces corresponding near real-time alert messages for dissemination via the VHF system to the world-wide scientific community interested in follow-up observations. The LBOT also produces messages to the on-board mass memory to secure the corresponding WFM data in full time and energy resolution from minutes before to minutes after the burst.
The LBOT is embedded inside the powerful FPGA at the core of the WFM Data Handling Unit (DHU). It is implemented in two ways:
i) A firmware part managing the LBOT interface with the rest of the DHU, the real-time pre-processing and binning of the data produced by the 10 WFM cameras and their storage into memory for usage by the trigger.
ii) A software part implementing the trigger algorithms, using the pre-processed data to seek for the appearance of a new unknown transient source, to localise it on the sky, and to generate the alert messages.

Two simultaneous trigger algorithms are implemented in the LBOT:
a) The count rate trigger searches in a first step for count rate increases over the estimated background in both cameras of each WFM camera pair, on several time scales, energy bands and detector zones. In case of an excess detection, in a second step detector plane image deconvolution using the mask patterns produces sky images, in which the excess is localised and compared to a catalogue of known sources. A detection of a new unknown source triggers the
alert message sequence.
b) The Òimage triggerÓ algorithm is dedicated to search for long duration transient
events (above typically 10 s). It is based on long exposure sky images built cyclically (without prior count rate increase) to search for unknown sources.

The LOFT Burst Onboard Trigger (LBOT) functionality will be implemented in the FPGA in a dedicated 2D FFT processor for image deconvolution based on standard IP core modules and thus will be an integral part of the DHU.
Current estimates of FPGA resource utilization indicates that less than 50\% (including margin) of the available logic units are used. Although not foreseen to be needed, this would allow for a dual core LEON3FT to be implemented.
The DHU is direct heritage from the Data Processing Unit (DPU) of the MXGS instrument of the ESA Atmosphere Space Interactions Monitor (ASIM) mission\cite{asim} currently under Phase C/D development at DTU Space. The processor board for the WFM DHU will, as far as possible, be a carbon copy of the processor board for the ASIM MXGS DPU. The accompanying Connector Board will be custom designed for the WFM DHU. I/O lines are abundant on the MXGS DPU board and enable full SpaceWire communication with OBDH and BEE and provide additional interfaces (LVDS, RS-422, analog I/O) for PDU control and monitoring.


\section{Conclusions and Outlook} 
\label{sec:conclusions}

Although the LOFT mission has not been down-selected for launch within the M3 call, a long ($>$2 years) assessment study has culminated with the delivery of more than 3000 pages of technical documentation. Most of the trade-offs have been closed, leading to a robust design. Prototypes of several subsystems for the LAD and the WFM have been constructed by the LOFT team during the study phase, including the detectors, collimators, ASICs, FEE, MBEE, PBEE and DHU in order to show the readiness of the technology and the feasibility of the concept.

Based on the ESA technical and programmatic review, on the scientific reports by the Advisory Structure and on the Q\&A session with the Astronomy Working Group and Fundamental Physics Advisory Group, the feedbacks for LOFT in M3 have all been very positive. In particular the technology was evaluated as mature. The high level of readiness and maturity of the mission and payload design, as well as the clean and solid assessment of the unique science case still make LOFT a very competitive mission with a compelling science case, and for this reason the LOFT Consortium Board has recently confirmed the intention to continue the development aiming now at the upcoming M4 ESA launch timescale (2026).

The work of the MSSL-UCL and Leicester groups is supported by the UK Space Agency. The work of SRON is funded by the Dutch national science foundation (NWO). The work of the group at the University of Geneva is supported by the Swiss Space Office. The Italian team is grateful for support by ASI (under contract I/021/12/0), INAF and INFN. The work of the IRAP group is supported by the French Space Agency. RH acknowledges GA CR grant 13-33324S and MSMT LH13065. The work of IAAT on LOFT is supported by the \textsl{Bundesministerium f\"ur Wirtschaft und Technologie} through the \textsl{Deutsches Zentrum f\"{u}r Luft- und Raumfahrt e.V. (DLR)} under the grant number FKZ 50 OO 1110. The authors would like to thank Dr.$~$Barry Giles, University of Tasmania, for helpful comments and discussions during the planning of the electronics concept.


\bibliography{tex}   
\bibliographystyle{spiebib}   

\end{document}